\documentclass{article}

\usepackage{PRIMEarxiv}

\usepackage[utf8]{inputenc} 
\usepackage[T1]{fontenc}    
\usepackage{hyperref}       
\usepackage{url}            
\usepackage{booktabs}       
\usepackage{amsfonts}       
\usepackage{nicefrac}       
\usepackage{microtype}      
\usepackage{lipsum}
\usepackage{fancyhdr}       
\usepackage{graphicx}       
\usepackage{amsmath}
\usepackage{ulem}
\usepackage{xcolor}
\graphicspath{{media/}}     
\usepackage{placeins}

\title{Experimental determination of tantalum L-shell fluorescence yields and Coster-Kronig transition probabilities}

\author{
  Nils Wauschkuhn, Katja Frenzel, Burkhard Beckhoff, Philipp Hönicke \\
  Physikalisch-Technische Bundesanstalt \\
  Abbestr. 2-12 \\
  10587 Berlin\\
  Germany\\
  \texttt{nils.wauschkuhn@ptb.de} \\
}

\begin{document}
\maketitle


\begin{abstract}
Using radiometrically calibrated instrumentation of Physikalisch-Technische Bundesanstalt, the L-shell fluorescence yields and Coster-Kronig factors of tantalum (including the uncertainty budget) were experimentally determined based on transmission and X-ray fluorescence experiments. The determined fluorescence yields ($\omega_{L3} = 0.247(12)$, $\omega_{L2} = 0.278(15)$, $\omega_{L1} = 0.157(12)$) were independently validated trough XRR-GIXRF experiments. Both, the  Coster-Kronig factors ($f_{23} = 0.123(84)$, $f_{13} = 0.328(152)$, $f_{12} = 0.14(11)$) as well as the fluorescence yields are in good agreement with the most established databases in the field of X-ray fluorescence.     
\end{abstract}


\section{Introduction}

The knowledge of atomic fundamental parameters (FP) such as the fluorescence yield, the photoionization cross section and the Coster-Kronig transition probabilities is of great importance for any quantitative analysis involving X-ray fluorescence (XRF). The majority of the available experimental and theoretical FP values for different elements were obtained more than forty years ago. For some chemical elements and some FPs, the tabulated data is based solely on interpolations as no experimental or theoretical data exists. Unfortunately, the uncertainties of most tabulated FP data often are not available or only estimated. As this is certainly an improvable situation, the International initiative on X-ray fundamental parameters \cite{FPI} and others are working on revisiting and updating FP databases with new experiments and calculations employing state-of-the-art techniques.\\ 
In this work, the tantalum L-shell fundamental parameters, namely the fluorescence yields and the Coster-Kronig factors, are being experimentally redetermined. Tantalum is a key element in microelectronics\cite{Shoki2003, Choi_2004}, solar industry\cite{Yang2018}, medicine and more. On the other hand, the availability of experimentally determined Ta-L shell FPs is rather scarce. The majority of the available experimental data is older than 30 years and the uncertainties as estimated for the most common tabulations\cite{Krause1979, T.Schoonjans2011} are only estimated.
In this work, we apply the reference-free XRF equipment of PTB\cite{Beckhoff2008} and dedicated transmission and fluorescence measurements\cite{M.Kolbe2012} using to revisit these parameters for tantalum.

\section{Experimental}
\label{sec:exp}
\subsection{Photon-in/photon-out experiment}
The experiments were performed at the wavelength-shifter beamline BAMline\cite{Goerner2001} at the BESSY II electron storage ring. This beamline provides hard monochromatic X-ray synchrotron radiation in the photon energy range from 5 keV up to 60 keV. Usually, the double crystal monochromator (DCM, with Si(111) crystals, d$E/E = ~0.2$\% between 8 and 50 keV) is used for applications comparable to the one in this study. The experiments were carried out using an in-house developed vacuum chamber\cite{M.Kolbe2005a} equipped with calibrated photodiodes and an energy-dispersive silicon drift detector (SDD) with experimentally determined response functions and radiometrically calibrated detection efficiency\cite{F.Scholze2009}. The sample was placed into the center of the chamber by means of an x-y scanning stage and the incident angle $\theta_{in}$ between the surface of the sample and the incoming beam was set to 45$^\circ$. As a sample, we have obtained a nominally 250 nm thick Ta deposition on a Si$_3$N$_4$-membrane. The membrane has a thickness of nominally 1000 nm. Furthermore, a blank Si$_3$N$_4$-membrane deposition was used to subtract the membrane contribution.

For both samples, transmission experiments were performed in the vicinity of the Ta-L absorption edges between 7 keV and 13 keV. In addition, the X-ray fluorescence emission from the coated sample was measured for photon energies ranging from about 10 keV to 13 keV. From these experiments, the Ta L-shell fluorescence yields and the Coster-Kronig factors can be determined as follows.

The procedure to determine L-shell fluorescence yields, as well as Coster-Kronig factors using physically calibrated instrumentation for reference-free X-ray spectrometry of PTB is already quite well established\cite{M.Kolbe2012, P.Hoenicke2014, M.Kolbe2015,Menesguen2018}. Here, Sherman’s equation\cite{Sherman1955} provides the basis for the calculation of fluorescence intensities of thin foils. It is a product of the incident monochromatic photon flux, a fluorescence production factor for a given shell $\sigma_S$, an instrumentation factor containing the solid angle of detection and the detection efficiency and the self-attenuation correction factor. This factor considers the attenuation of the photons on their way through the sample: For the incoming photons $\Phi_0(E_0)$ the attenuation on their way to the point of interaction is considered, for the fluorescence photons $\Phi^d_i(E_0)$ the attenuation on their way from the point of interaction to the detector is considered. Employing tunable photon sources or as recently shown also employing energy dispersive detectors\cite{Huang_2021}, this factor can be easily determined by transmission measurements for the relevant photon energies.

The fluorescence production factor $\sigma_{Li}$ is defined as follows: 
\begin{align}
\sigma_{L3}(E_0) &= \omega_{L3} (\tau_{L3}(E_0) + f_{23}\tau_{L2}(E_0) + [f_{13} +  f_{12} f_{23}] \tau_{L1}(E_0)) \\
\sigma_{L2}(E_0) &= \omega_{L2} (\tau_{L2}(E_0) + f_{12}\tau_{L1}(E_0)) \\
\sigma_{L1}(E_0) &= \omega_{L1} \tau_{L1}(E_0)
\end{align}
It is depending on the photon energy $E_0$ and is calculated employing the respective subshell fluorescence yield $\omega_{Li}$, the subshell photoionization cross sections $\tau_{Li}(E_0)$ as well as the Coster-Kronig factors $f_{ji}$. The latter are irrelevant for photon energies below the edge energy of the respective subshell as the photoelectric cross section is zero for energies below the corresponding subshell threshold energy. Thus, for photon energies between $E_{L3}$ and $E_{L2}$, $\sigma_{L3}(E_0)$ is simply the product of fluorescence yield and photoionization cross section so that the fluorescence yield $\omega_{L3}$ can be derived. By further employing this selective excitation to the other edges, also the $L_2$ and $L_1$ subshell fluorescence yields as well as the Coster-Kronig factors can be determined.

\begin{figure}
  \centering
    \includegraphics[width=13cm]{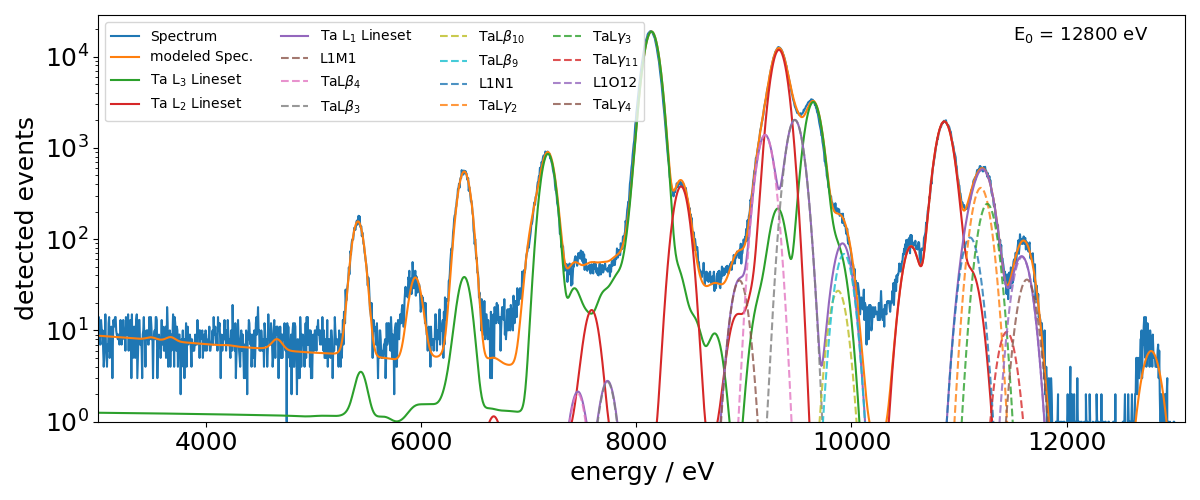}
  \caption{Exemplary fluorescence spectrum recorded on the Ta film at $E_0 = 12.8$ keV in blue together with the overall deconvolution (orange) as well as selected response functions for the fixed $L_i$ linesets. For comparison, the single fluorescence lines of the $L_1$ lineset are also plotted as dashed lines.}
  \label{fig:fig1}
\end{figure}

In other words, if $E_{L3} \leq E_0 \le E_{L2}$, the fluorescence production factor for $L_3$ reduces to

\begin{equation}
\sigma_{L3}(E_0)\rho d = \omega_{L3} \tau_{L3}(E_0)\rho d = \frac{\Phi^d_i(E_0)M_{i,E_0}}{\Phi_0(E_0)\frac{\Omega}{4\pi}}
\label{eq:prodCS}
\end{equation}
with
\begin{equation}
M_{i,E_0} = \frac{(\frac{\mu_S(E_0)\rho d}{\sin \theta_{in}}+\frac{\mu_S(E_i)\rho d}{\sin \theta_{out}})}{(1-\exp[-(\frac{\mu_S(E_0)\rho d}{\sin \theta_{in}}+\frac{\mu_S(E_i)\rho d}{\sin \theta_{out}})])},
\label{eq:M}
\end{equation}
where $\theta_{in}$ and $\theta_{out}$ are incident and exit angles respectively.
Due to the use of PTB's physically calibrated instrumentation for reference-free X-ray spectrometry, all of the relevant measures can be accessed. The fluorescence photon flux $\Phi^d_i(E_0)$ is derived from the recorded fluorescence spectra by means of a spectral deconvolution procedure. Here, the detector response functions for all relevant fluorescence lines as well as relevant background contributions, e.g. bremsstrahlung, originating from photo-electrons are included. In addition, we determine and apply fixed line sets for each of the three L-shells in order to stabilize the deconvolution\cite{M.Kolbe2012}. An exemplary spectrum including the deconvolution is shown in fig. \ref{fig:fig1}. The incident photon flux $\Phi_0(E_0)$ and the solid angle of detection $\frac{\Omega}{4\pi}$ are known due to the use of calibrated instrumentation \cite{Beckhoff2008}. The sample specific attenuation correction factor $M_{i,E_0}$ for the incident ($E_0$) -- as well as the fluorescence radiation ($E_i$) is calculated according to Eq. \ref{eq:M} using the experimentally determined sample specific attenuation coefficients $\mu_S(E_0)\rho d$ and $\mu_S(E_i) \rho d$.

\begin{figure}
  \centering
    \includegraphics[width=8cm]{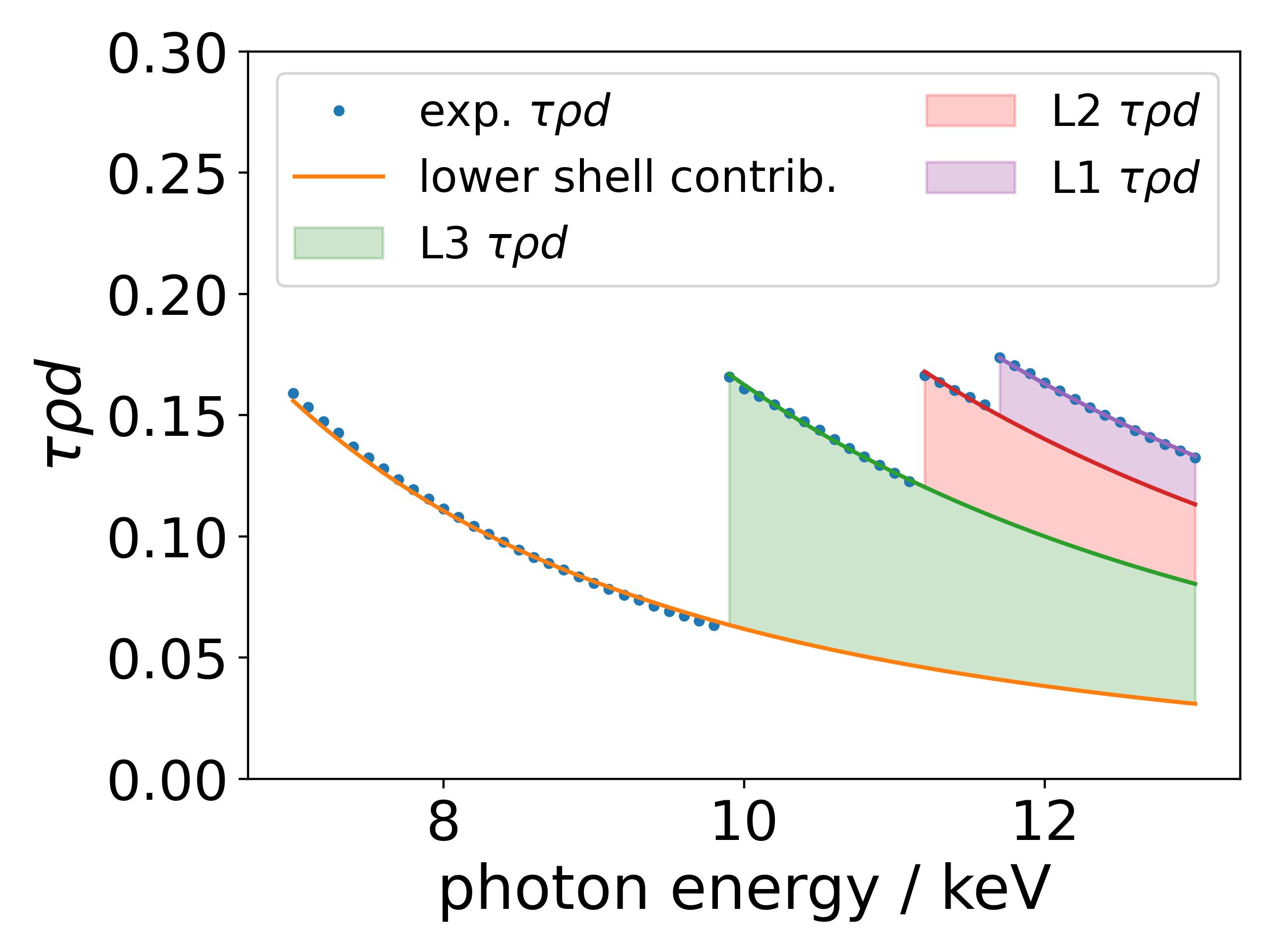}
  \caption{Experimentally determined $\tau(E_0)\rho d$ (blue dots) for the employed tantalum thin film and its separation into the
lower bound shells (orange line) as well as the L$_3$ (green), L$_2$ (red) and L$_1$ (purple) contributions.
}
  \label{fig:fig2}
\end{figure}

Employing the experimental $\mu_S(E_0)\rho d$ and $\mu_S(E_i) \rho d$ values, one can calculate the total sample specific photoionization cross sections $\tau_S(E_0)\rho d$ and $\mu_S(E_i) \rho d$ by removing the scattering contributions. For this purpose, we derive the relative scattering contribution at each photon energy from a database (e.g. X-raylib) and use this data to determine $\tau_S(E_0)\rho d$. The thereby obtained photoionization cross sections are shown in figure \ref{fig:fig2} as blue dots. For the determination of the subshell fluorescence yields as well as the Coster-Kronig factors, one needs to isolate the subshell photoionization cross sections $\tau_{Li}(E_0)\rho d$. This is performed by scaling the Ebel polynomials \cite{H.Ebel2003} for the lower bound shells as well as the three L-subshells to the $\tau_S(E_0)\rho d$ as shown in the figure. 

\begin{figure}
  \centering
    \includegraphics[width=7.5cm]{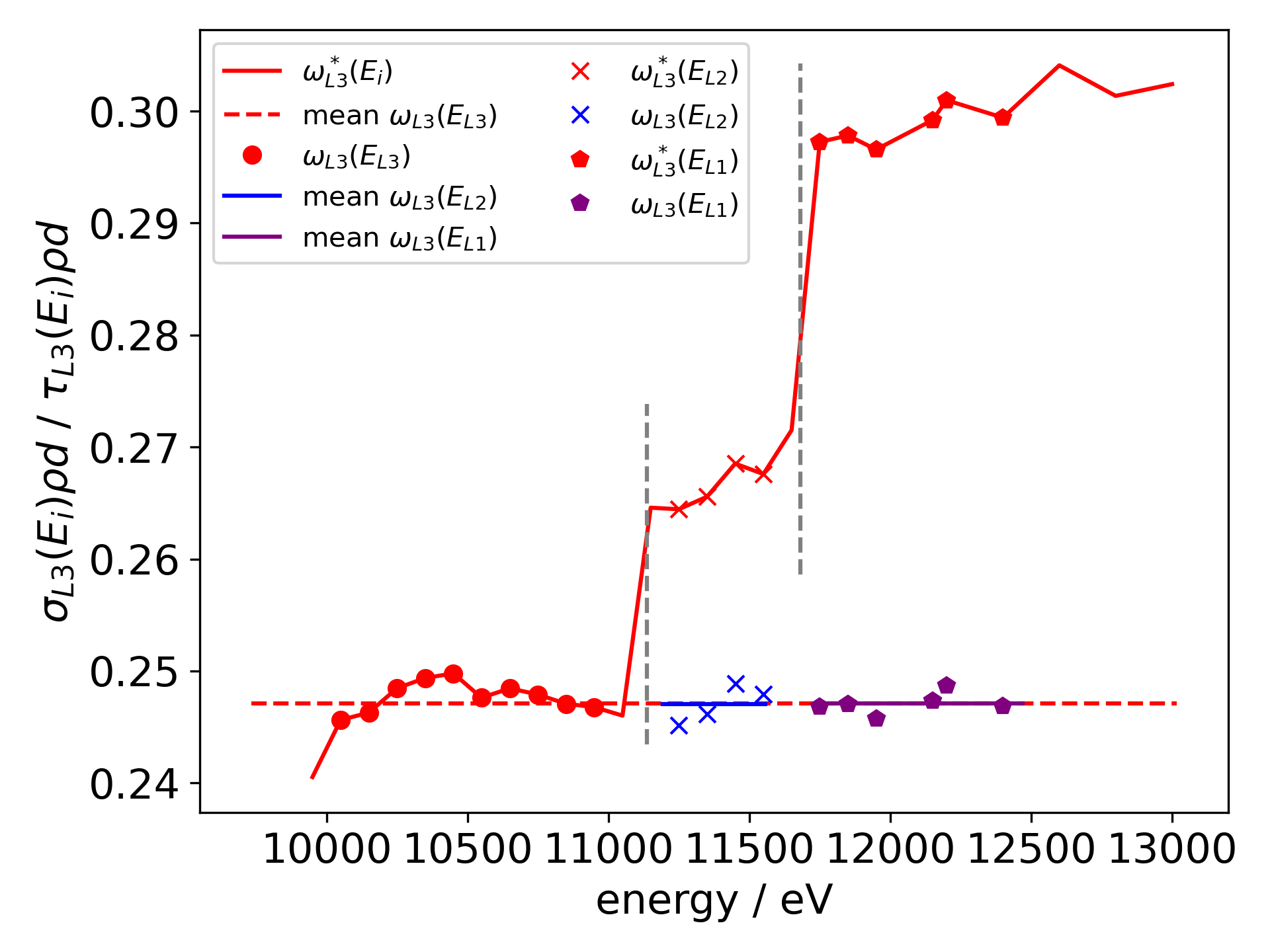}
    \includegraphics[width=7.5cm]{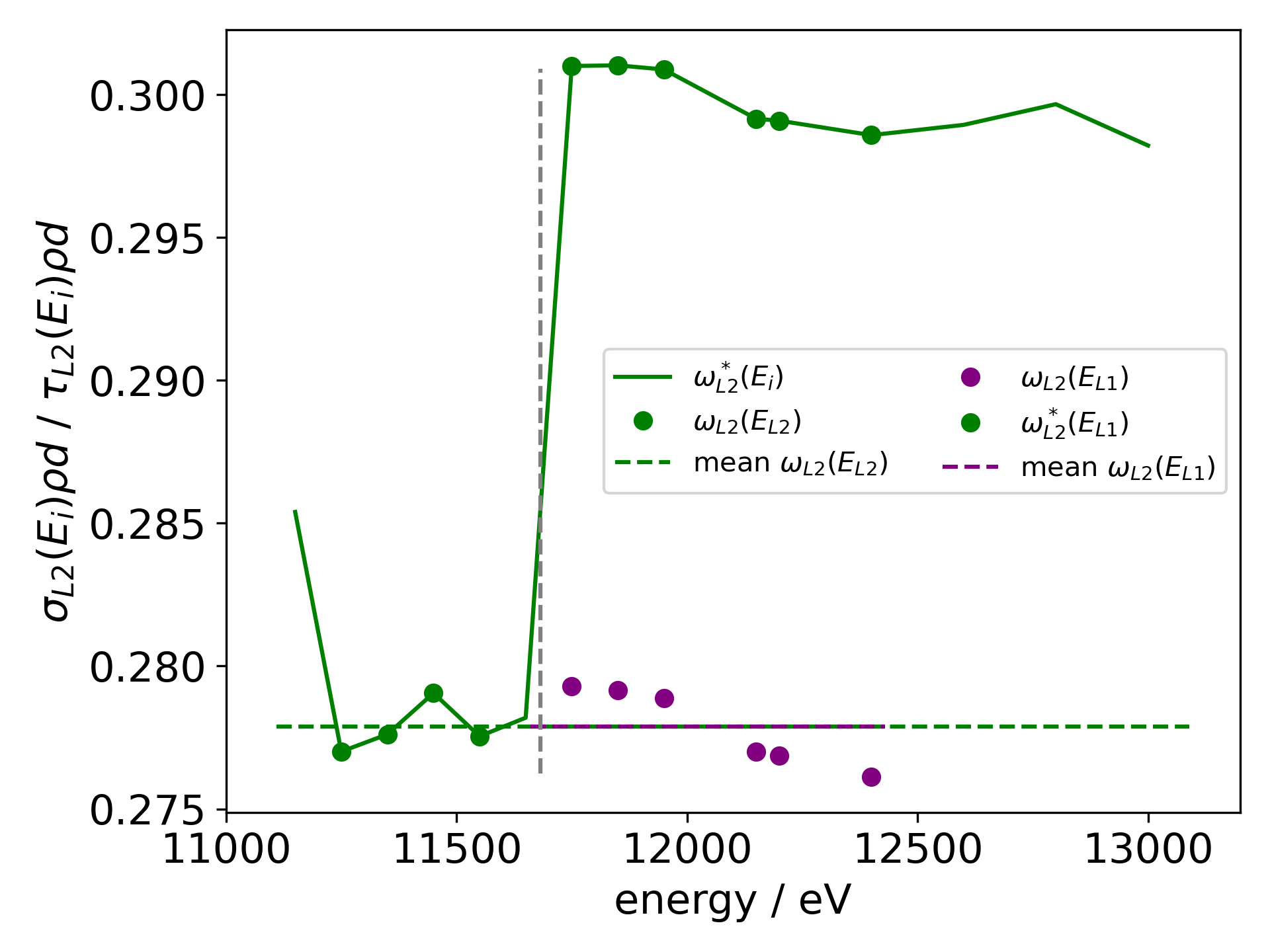}
  \caption{Experimentally determined Ta-L$_3$ (left image) and Ta-L$_2$ (right image) fluorescence yield versus excitation photon energy without taking into account the Coster-Kronig transitions (red or green symbols), as well as the mean value for CK transitions turned off (horizontal dashed lines). The CK factors are chosen in order to match the average corrected fluorescence yields with the dashed lines (blue and purple symbols and lines). The vertical dashed lines mark the L$_2$ and L$_1$ absorption edge above which the respective CK transition appears.}
  \label{fig:CKs}
\end{figure}

With the $\tau_{Li}(E_0)\rho d$, all relevant parts of eq. \ref{eq:prodCS} are known and it can be solved for the fluorescence yield. The same procedure is applied for the $L_2$ and $L_1$ shells. By applying the same procedure for incident energies above the subsequent absorption edges, the Coster-Kronig factors can be derived. In figure \ref{fig:CKs}, this is shown for the case of $\omega_{L3}$ and $\omega_{L2}$. Here, the derived fluorescence yield values marked with a star, for example $\omega^*_{L3}(E_{i})$ are being calculated by only taking into account the normalized fluorescence intensity of the $L_3$ shell as well as the derived $\tau_{L3}(E_i)\rho d$ (red line in figure \ref{fig:CKs}). If the incident photon energy is above the subsequent absorption edge (marked as grey dashed vertical lines), the $\omega^*_{L3}$ jumps due to the additional Coster-Kronig related contributions to the total effective photoionization cross section, namely the term $f_{23}\tau_{L2}(E_0)$ in the case shown in figure \ref{fig:CKs} as red crosses. As the fluorescence yield value must be constant and not dependent on the incident photon energy, the Coster-Kronig factor can be determined so that the Coster-Kronig corrected $\omega_{L3}(E_{L2})$ matches the one determined for an excitation below the $L_2$ absorption edge.  

\subsection{Validation measurement with combined XRR and GIXRF}
\begin{figure}
  \centering
    \includegraphics[width=9cm]{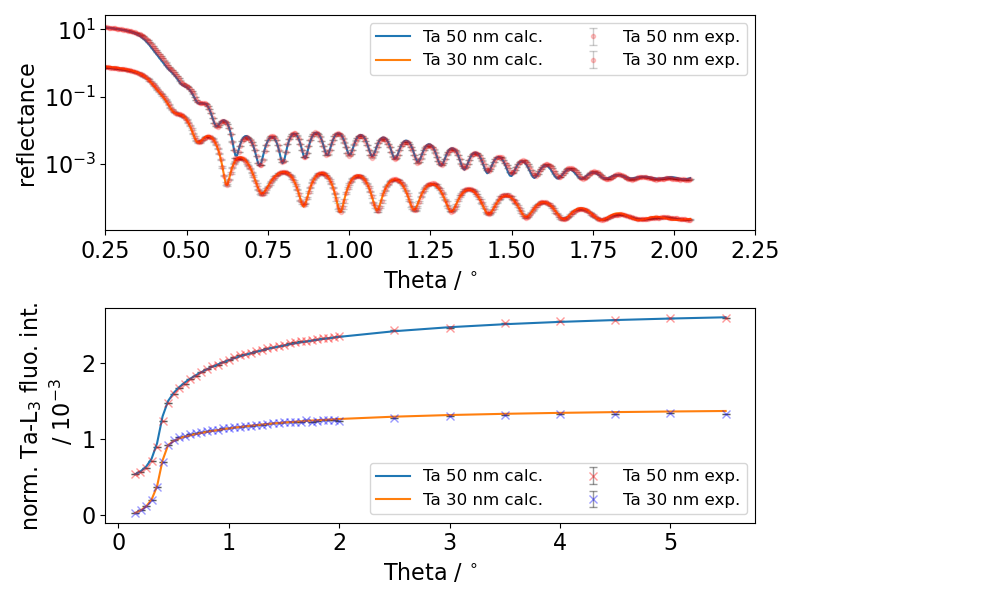}
  \caption{Comparison of measured and calculated data: reflectance (top) and the normalized fluorescence intensity (bottom)}
  \label{fig:GIXRF-XRR}
\end{figure}

As an independent validation of the experimentally determined L$_3$ fluorescence yield, a combined reference-free GIXRF-XRR measurement {\cite{Hoenicke_2019}} was carried out on two different tantalum layer samples. These experiments have been carried out employing an in-house build ultrahigh vacuum chamber dedicated to reference-free XRS\cite{J.Lubeck2013} at PTB’s four-crystal-monochromator (FCM) beamline \cite{Krumrey1998}. The two layer samples employed consist of pure Ta layers on silicon wafers with nominal thicknesses of 30 nm and 50 nm. 
The energy of the incident beam was set to 10\,keV to only excite the L3 shell of tantalum. For both samples, the incident angle dependent Ta-L$_3$ fluorescence emission as well as the reflected incident radiation have been recorded. The experimental data including a basic evaluation (spectra deconvolution, normalization to incident photon flux and solid angle of detection) are shown in Fig. \ref{fig:GIXRF-XRR}.\\

To determine the FPCS (fluorescence production cross section) from the experimental data, a quantitative combined modeling of the GIXRF-XRR data was performed as shown in ref.\cite{Hoenicke2022}. For this purpose, a model based on a thin carbonaceous contamination layer on tantalum oxide on tantalum on native oxide covered silicon was used. For each layer, with the exception of the substrate, the thickness and relative density were used as model parameters. In addition, the top surface roughness as well as the tantalum layer roughness were modeled. The roughness of the tantalum oxide layer was set to be the same as the previous. Experimental parameters such as the beam divergence or the photodiode's dark current were modeled as well. The modeling process is realized using the Sherman equation \cite{Sherman1955} which is shown below, for the GIXRF measurement and using the matrix method \cite{Abel_s_1950} for the XRR measurement.

\begin{align}
\frac{4\pi\sin\theta_i}{\Omega(\theta_i)}\frac{F(\theta_i,E_i)}{\Phi_0\epsilon_{E_f}} &= W_i \rho \tau(E_i) \omega_{L3} dz \cdot \sum_{z} P(z) \cdot I_{XSW}(\theta_i,E_i,z) \cdot \exp\left[-\rho\mu_{E_f}z\right]\textrm{.}
\label{eq:sherman}
\end{align}

Here, the experimentally derived fluorescence count rate $F(\theta_i,E_i)$ of the lineset related to the Ta-L$_3$-edge, excited using photons of energy $E_i$ at an incident angle $\theta_i$ is the essential measurand. A normalization on the effective solid angle of detection $\frac{\Omega(\theta_i)}{4\pi}$, the incident photon flux $\Phi_0$, and the detection efficiency of the used fluorescence detector $\epsilon_{E_f}$ is also required. By calculating the X-ray standing wave field intensity distribution $I_{XSW}(\theta_i,E_i,z)$, a numerical integration in conjunction with the depth distribution $P(z)$ of the tantalum distribution and an attenuation correction factor, the experimental data can be reproduced. For a quantitative modeling, the atomic fundamental parameters, namely the L$_3$-subshell photoionization cross section $\tau(E_i)$ and the fluorescence yield $\omega_{L3}$, and material-dependent parameters, e.g. the weight fraction $W_i$ of element $i$ within the matrix as well as the density $\rho$ of the matrix must also be considered. For the latter, we have adopted the density of the 50 nm Ta layer (14.2 $\frac{g}{cm^3}$) from a previous study of the same sample\cite{Ciesielski_2022} in order to reduce the degrees of freedom. The ratio of this density with respect to the Ta bulk density was applied to both the Ta$_2$O$_5$ and the Ta layers of both samples.

The relevant optical constants were taken from X-raylib \cite{T.Schoonjans2011} using the respective $\rho_{bulk}$ and are also scaled using each material's relative density. The FPCS for the Ta-L$_3$ shell was also taken from X-raylib and is scaled employing a factor during the modeling. The optimization was performed using a Markov chain Monte Carlo (MCMC) algorithm \cite{Foreman_Mackey_2013}.

The final model calculations are also shown in Fig. \ref{fig:GIXRF-XRR} and agree very well with the experimental data. The determined layer thickness of the Ta layers is about 28.9 nm or 46.9 nm and thus reasonably in line with the nominal values. 
\section{Results}
\begin{figure}
  \centering
    \includegraphics[width=14.8cm]{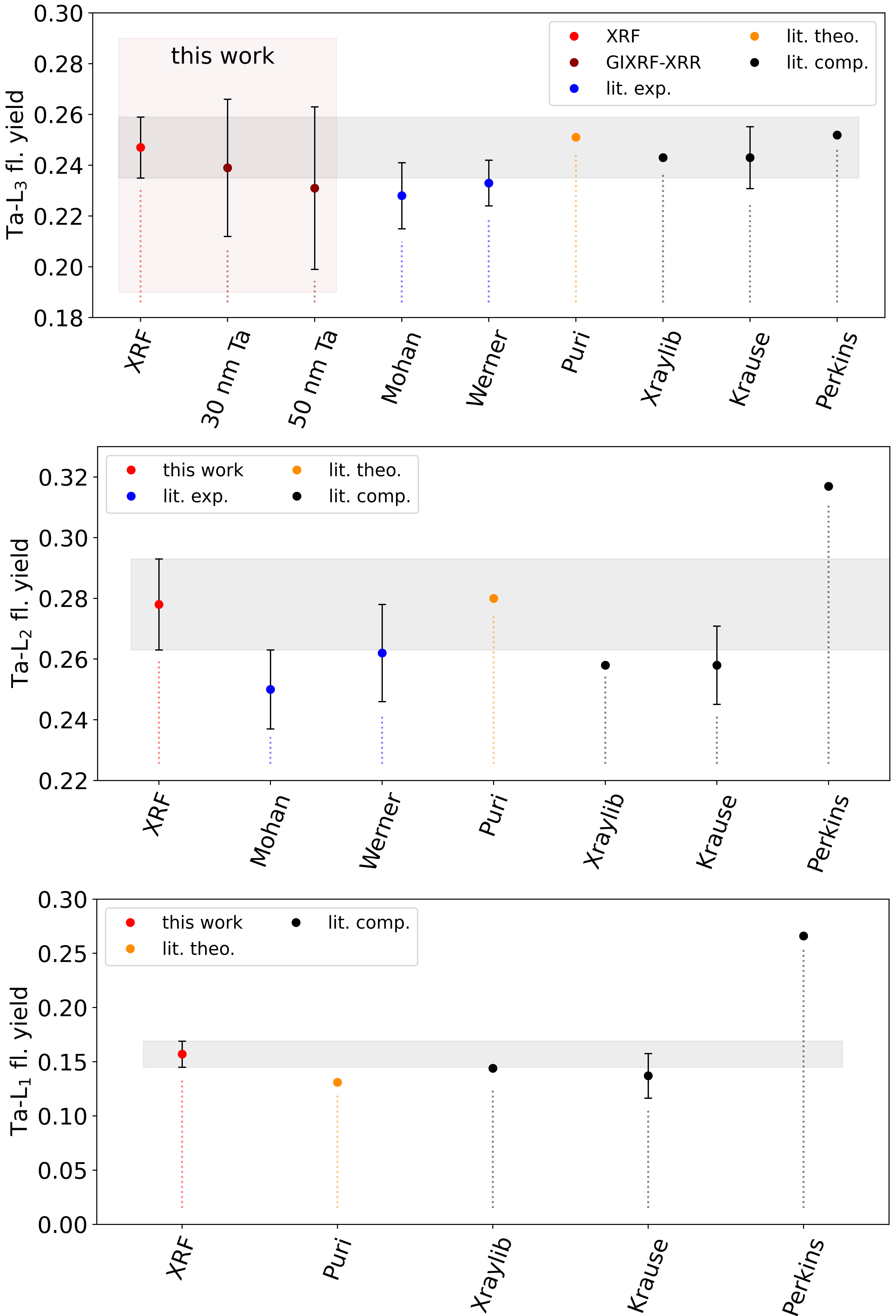}
  \caption{Experimentally determined Ta-L fluorescence yields in comparison to selected literature data from other experimental works of Mohan\cite{Mohan_1970} and Werner\cite{Werner_1988} (blue), theoretical calculations by Puri \cite{S.Puri1993} (orange) or commonly used compilations\cite{Krause1979, T.Schoonjans2011, S.T.Perkins1991} (black). The experimental uncertainties of our values are plotted as grey boxes for easier comparison. For the Ta-L$_3$ fl. yield, both the XRF result and the GIXRF-XRR values are shown.}
  \label{fig:YieldL3}
\end{figure}

The results derived for the Ta-L subshell fluorescence yields are shown in figure \ref{fig:YieldL3} as well as in table \ref{tab:table} in comparison to selected data from the literature. They were averaged from the values derived at the different excitation photon energies below the subsequent absorption edge as indicated in figure \ref{fig:CKs}. The uncertainty budget of the determined fluorescence yields is calculated using the relative uncertainty contributions of the involved parameters. The main contributors to the total uncertainty budget are the determined subshell photoionization cross sections (~2.5 \% for L$_3$, ~6 \% for L$_1$) and the uncertainty contribution of the spectral deconvolution (~2 \%). The uncertainty budget one can achieve by employing PTB's reference-free XRF approach for the determination of atomic fundamental parameters is discussed in more detail in ref. \cite{Unterumsberger2018}.

In general, our experimental values agree reasonably well with commonly used X-raylib\cite{T.Schoonjans2011} data and the theoretical predictions of Puri et al.\cite{S.Puri1993}. Significant deviations larger than the uncertainty are observed for the L$_2$-shell yield with respect to X-raylib and the L$_1$-shell yield with respect to Puri. The agreement to the shown older experimental data is good considering the stated respective uncertainties. It should be noted, that there are more published values for the L-subshell fluorescence yields of Ta with different origins (experimental or interpolated). They are nicely summarized in a recent work by Sahnoune et al.\cite{Y.Sahnoune2016}.

From the GIXRF-XRR modeling, Ta-L$_3$ fluorescence yield values of 0.239(27) (sample A, 30nm Ta) and 0.231(32) (sample B, 50nm Ta) assuming that the L$_3$-subshell photoionization cross section for Ta from X-raylib at the employed excitation photon energy of 10 keV is correct. These results are also shown in tab. \ref{tab:table} and figure \ref{fig:YieldL3} in combination with the other data. The uncertainty of the GIXRF determined value is estimated based on the confidence interval of the modeling and an estimated uncertainty of the tabulated L$_3$-subshell photoionization cross section. Unfortunately, the uncertainties are too large in order to reliably judge which fluorescence yield is more accurate. For such small deviations between the determined experimental value and the tabulated value, the sensitivity of the GIXRF-XRR approach is not sufficient. This is mainly due to the strong parameter correlation with the layer densities. If the densities could be determined independently and thus kept fixed for the modeling, it would significantly improve the sensitivity for the fluorescence production cross section.

\begin{figure}
  \centering
    \includegraphics[width=13cm]{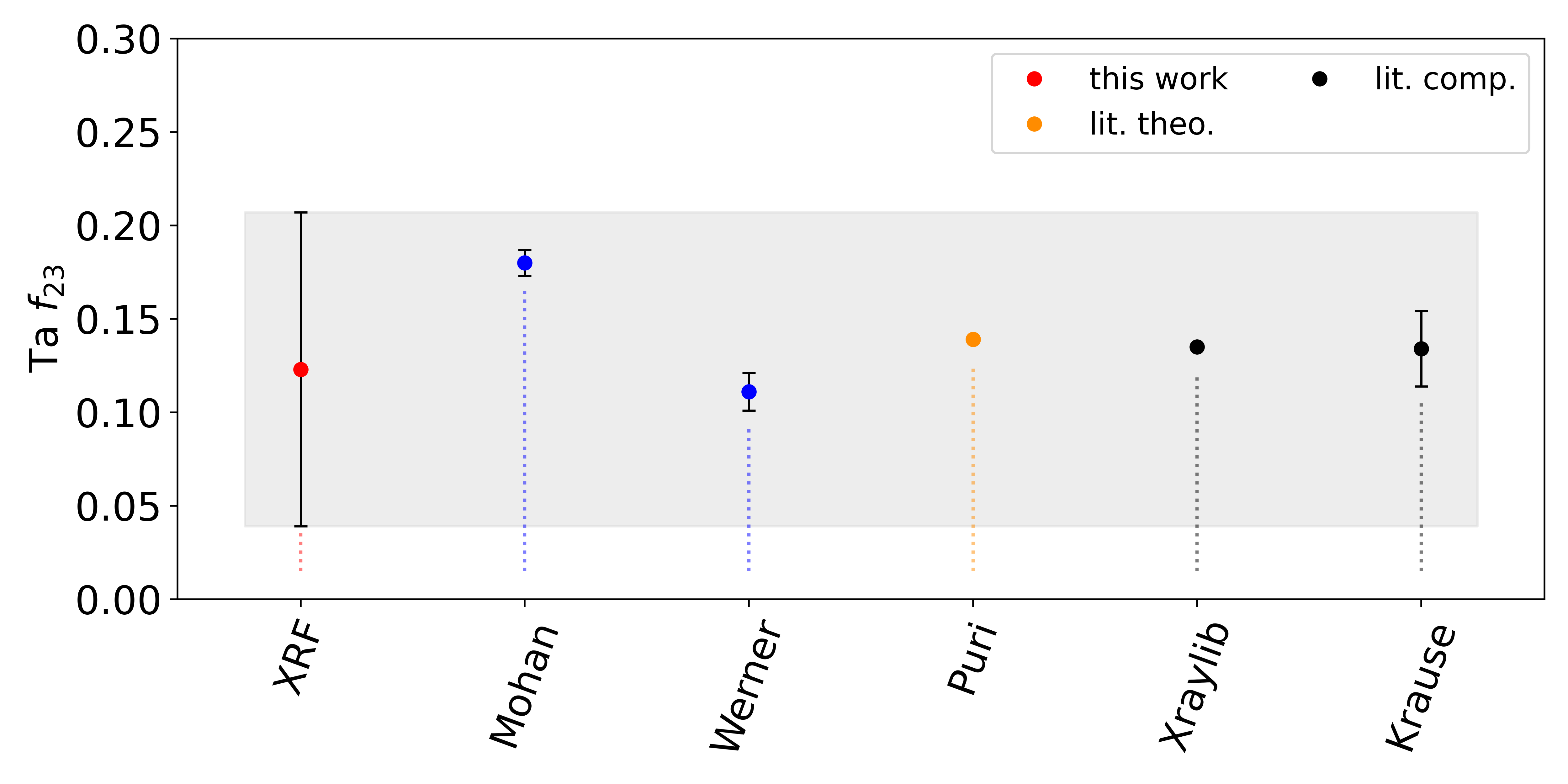}
  \caption{Experimentally determined Coster-Kronig factor $f_{23}$ for Ta-L fluorescence in comparison to selected literature data from other experimental works of Mohan\cite{Mohan_1970} and Werner\cite{Werner_1988} (blue), theoretical calculations by Puri \cite{S.Puri1993} (orange) or commonly used compilations\cite{Krause1979, T.Schoonjans2011} (black). The experimental uncertainty of our value is plotted as a grey box for easier comparison.}
  \label{fig:CK23}
\end{figure}

The experimentally determined Coster-Kronig factors are also shown in table \ref{tab:table} in comparison to selected data from the literature. For $f_{23}$ also a graphical comparison is shown in figure \ref{fig:CK23}. The derived value for $f_{23}$ is in good agreement with the data from the literature, even when considering a much lower uncertainty budget. A similar behaviour can be found for $f_{13}$, where the observed differences are much lower than our stated uncertainty. Only for $f_{12}$, the deviations with respect to the commonly used database values are somewhat large but still within our stated uncertainty.

The determined relative uncertainties of the Coster-Kronig factors are higher than those of the fluorescence yields because of the required error propagation. For the Coster-Kronig factors, the relation between the different subshell photoionization cross sections increases the total relative uncertainty. Hence, a reliable uncertainty budget for
the determined Coster-Kronig factors leads to large uncertainties in the order of the values itself\cite{M.Kolbe2012}. But as can be seen by the non-agreeing values by Mohan and Werner, our uncertainty seems more reasonable and is more reliable.

\begin{table}
 \caption{Overview of the experimentally determined Ta-L subshell fluorescence yields and Coster-Kronig factors as well as a comparison to the most commonly used database \cite{T.Schoonjans2011} and selected values from the literature. For the GIXRF-XRR results, the value marked with A refers to the 30 nm Ta sample and the other to the 50 nm sample.}
  \centering
  \begin{tabular}{c|c|c|c}
    \toprule
    \toprule
    & Ta $\omega_{L3}$     & Ta $\omega_{L2}$     & Ta $\omega_{L1}$ \\
    \midrule
    this work (XRF) & 0.247(12) & 0.278(15)  & 0.157(12)     \\
    this work (GIXRF-XRR) & 0.239(27) (A), 0.231(32) (B) & & \\
    X-raylib \cite{T.Schoonjans2011} & 0.243 & 0.258  & 0.144     \\
    Puri et al. \cite{S.Puri1993} & 0.251 & 0.28  & 0.131     \\
    Werner et al. \cite{Werner_1988} & 0.233(9) & 0.262(15)  &  \\
    \toprule
    \toprule
   &    Ta $f_{23}$     & Ta $f_{13}$     & Ta $f_{12}$ \\
    \midrule
       this work (XRF) &  0.123(84)     & 0.328(152)       & 0.14(11)  \\
    X-raylib \cite{T.Schoonjans2011}  &  0.135     & 0.351       & 0.186  \\
    Puri et al. \cite{S.Puri1993} & 0.139 & 0.351  & 0.186     \\
    Werner et al. \cite{Werner_1988} & 0.111(10) & 0.339(20)  & 0.104(15) \\
   \bottomrule
  \end{tabular}
  \label{tab:table}
\end{table}
\FloatBarrier
\section{Conclusion}
The tantalum L-shell fluorescence yields and Coster-Kronig factors have been experimentally determined employing the radiometrically calibrated instrumentation of PTB using Ta coated Si$_3$N$_4$-membranes. The determined fluorescence yields are agreeing well with the commonly used X-raylib tables except for the L$_2$-shell. Here, our value is slightly larger than the tabulated value. The achieved experimental uncertainties for the three fluorescence yields are in the same order as the Krause estimates\cite{Krause1979}. This not only puts the estimated uncertainties on more solid grounds, it also allows to conclude with reasonable reliability that the estimated uncertainties for the L-shell yields of neighbouring elements are in the right order of magnitude as well.
Considering both the determined fluorescence yields and the Coster Kronig factors, we can conclude that the X-raylib table gives a relatively good collection of the relevant Ta-FPs. This is also in line with observations from earlier FP determinations from our group\cite{Hoenicke2022, Unterumsberger2018, P.Hoenicke2016a}. Thus, the X-raylib tables are a very good starting point if a consistent database is needed.

\FloatBarrier
\section*{Acknowledgments}
This project has received funding from the ECSEL Joint Undertaking (JU) IT2 under grant agreement No 875999. The JU receives support from the European Union’s Horizon 2020 research and innovation programme and the Netherlands, Belgium, Germany, France, Austria, Hungary, United Kingdom, Romania and Israel.

\bibliographystyle{unsrt}  
\bibliography{references}

\end{document}